\documentclass{mem}
\usepackage{natbib}\usepackage{txfonts}\usepackage{balance}
\usepackage{graphicx}
\usepackage[a4paper]{hyperref}
\idline{75}{282}
\begin{document}
\def\teff{$T\rm_{eff }$}
\def\kms{$\mathrm {km s}^{-1}$}
\def\gsimeq{\hbox{\raise0.5ex\hbox{$>\lower1.06ex\hbox{$\kern-1.07em{\sim}$}$}}}

\title{
Identification of (high-redshift) AGN with WFXT: 
lessons from COSMOS and CDFS}

   \subtitle{}

\author{
M. \,Brusa\inst{1} 
\and R. \, Gilli\inst{2}
\and F. \, Civano\inst{3}
\and A. \, Comastri\inst{2}
\and F. \, Fiore\inst{4}
\and C. \, Vignali\inst{5}}

  \offprints{M. Brusa}

\institute{
Max Planck Institut f\"ur Extraterrestrische Physik,
Giessenbachstrasse 1, D-85748 Garching by M\"unchen, Germany 
\email{marcella@mpe.mpg.de}
\and  
Istituto Nazionale di Astrofisica --
Osservatorio Astronomico di Bologna, Via Ranzani 1,
I-40127 Bologna, Italy
\and
Harvard-Smithsonian Center for Astrophysics, 60 Garden Street, Cambridge, MA 02138 
\and
Istituto Nazionale di Astrofisica --
Osservatorio Astronomico di Monteporzio Catone, Via Frascati 33,
I-00044 Monte Porzio Catone, Italy
\and 
Dipartimento di Astronomia -- Universit\`a di Bologna, 
Via Ranzani 1, I-40127 Bologna, Italy
}

\authorrunning{Brusa }

\titlerunning{Identifications in WFXT}

\abstract{The Wide Field X--ray Telescope (WFXT) will provide tens of millions of
AGN, with more than 4$\times10^5$ expected at z$>3$.
Here we review the issues present in the identification of (large) samples of faint
and high-redshift X--ray sources, and describe a statistical, powerful tool that can be applied
to WFXT catalogs. The depth of associated optical and near infrared catalogs, 
needed for a reliable and as much complete as possible identification, are also discussed,
along with the combined synergies with existing or planned facilities. 


}
\maketitle{}

\section{Scientific drivers}

One of the main aims in extragalactic astronomy for the next decade 
is the study of the co-evolution of galaxies and the Super Massive
Black Holes (SMBH) residing in their centre, out to the very
first epochs of galaxy formation. In this respect, deep and sensitive 
X--ray observations will be the {\it unique} instrument to reveal the 
accretion light from SMBH in galactic nuclei at high-z, which are often 
invisible at longer wavelengths because of intergalactic absorption and 
dilution by the host galaxy. 

%Per sottolineare l'AGN demography con il WFXT, direi che al momento studi ad 
%alto z sono prevalentemente su oggetti broad-line e brillanti (SDSS-like), 
%per cui l'X ha permesso di studiarne le proprieta' di accrescimento, almeno 
%in termini di proprieta' medie. Se invece si va sulla pop. di X-ray selected, 
%allora finora le surveys sono state poco efficienti e ce ne sono tra 10 e 20 
%a z>4, e qui di seguito quello che dici con "present X-ray surves ..."

The study of Active Galactic Nuclei (AGN) demography at z$>3$ 
is one of the key science drivers for the Wide Field X--ray Telescope 
(WFXT, e.g. \citealt{forman}). 
In the past decade, the characterization of the early phase of SMBH growth has
been limited to the study of optically selected QSOs detected mostly
in the SDSS survey, i.e. sampling only the unobscured and most luminous
tail of the AGN population. 
Deep and medium deep {\it Chandra} and XMM-{\it Newton} surveys have allowed the study
of X--ray selected QSOs up to relatively high redshifts, z$\sim3-4$. 
At higher redshifts, present X--ray
surveys are highly incomplete and strongly limited by the small area sampled. 
As an example, there are only a few X--ray selected QSO with confirmed spectroscopic 
redshifts at z$>5$ \citep[see][]{barger} %,
%and a few more candidates from deep X--ray surveys for which only photometric 
%redshifts are available \citep[e.g., in the CDFS][]{luo10}. 
Moreover, the extrapolations of the X--ray luminosity function 
(LF) as obtained combining various XMM and Chandra surveys 
differ by up an order of magnitude (see Figure 1, and reference therein, 
adapted from \citealt{b10}). 
As a comparison, the number of optically selected QSOs revealed up to z$\sim6$ is approaching 
50, i.e. large enough to determine their LF which encodes the information about 
the history of SMBH build up and the integrated flux of UV ionizing radiation \citep[e.g.][and reference therein]{fan,jiang,willott}.
%The space density and LF of X--ray selected luminous (logL$_{\rm X}>44$ QSO) at high redshifts (z$>4$) nowadays 
%remains essentially unconstrained.
An unbiased search of X--ray selected z$\sim5-6$ QSOs would require to survey several
hundreds of square degrees to a depth of the order of $\sim10^{-15}$ erg cm$^{-2}$ s$^{-1}$ 
and thus beyond the capabilities of current X--ray telescopes. 
WFXT will offer the unique opportunity to explore 
the high-redshift universe, providing {\it about two order of magnitudes} larger samples with respect 
to the current SDSS samples ($\sim2000$ z$>6$ AGN vs. $\sim50$), 
opening a completely new, unexplored window for LF analysis.
%Comparison of Chandra/XMM works and SDSS optical selection: needs a two/three 
%order of magnitudes larger statistics to match the samples (X--rays needed to 
%get the faint end of LF, more representative of the whole high-z
%population rather than SDSS)
%More specifically, at the bright fluxes ($>5\times10^{-16}$ cgs) data and predictions robust,
%200 deg$^2$ needed to have same statistics of SDSS at COSMOS depth.  
%At low fluxes ($<10^{-16}$ cgs), data scarse, predictions uncertain 
%(CDFS analysis predicts a factor of 2 higher than extrapolations from bright 
%fluxes). 
%The expectations at different redshifts and in the three different WFXT 
%surveys (REFE) are reported in Table 1. In the deep survey, and for the
%highest redshift bin, the estimates have been made both assuming the most conservative 
%approach based on extrapolating the Gilli et al. (2007) model which best fit the low-z data,
%and assuming a factor two higher surface density as suggested by the CDFS analysis 
%(see Fiore et al., this book).
(see Gilli et al. 2010, this volume, for a full description of the 
high-redshift AGN demography with WFXT).

\section{Identification issues}
The identification of the correct counterparts of both obscured 
and unobscured AGN is the first, crucial step for a full characterization 
of the physical and evolutionary properties of the entire population. 
At high redshifts, this process is further complicated by the fact 
that 1) z$>3$ sources constitute only a tiny fraction ($\sim1$\%)
of the entire X--ray population  ($<0.1$\% for z$>6$ sources) and 
2) these objects are usually faint in the optical band, because 
the emission would be strongly reduced by cosmological dimming,
and/or, for obscured sources, the intrinsic AGN emission is absorbed by 
the surrounding material. 
As a result, the probability of finding by chance a galaxy of R$>24$ in 
the X--ray error box is not negligible even with Chandra given the 
high surface density of background galaxies
\cite[see extensive discussions in, e.g.,][]{luo10}.
%
%circle depends on the accuracy of the X--ray source positioning and on
%the magnitude of the optical counterpart. 
%At R$>24$ the probability of chance association 
%
The identification process is made easier by using deep near infrared 
images 
%the surface density of these sources being smaller than that 
%of optical ones.
%bringing the probability of finding a galaxy by chance 
%in $\sim2"$ radius error boxes to comfortably small values.  
%Moreover, 
given that AGN are strong IR emitters and 
the K-band flux is more tightly correlated with the X-ray 
flux than the optical (obscured) one 
%and it can be considered a better driver for the identification 
\citep[see][]{b09b}.

%%%% Two columns %%%%%%%
\begin{figure}[t!]
\resizebox{\hsize}{!}{\includegraphics[clip=true]{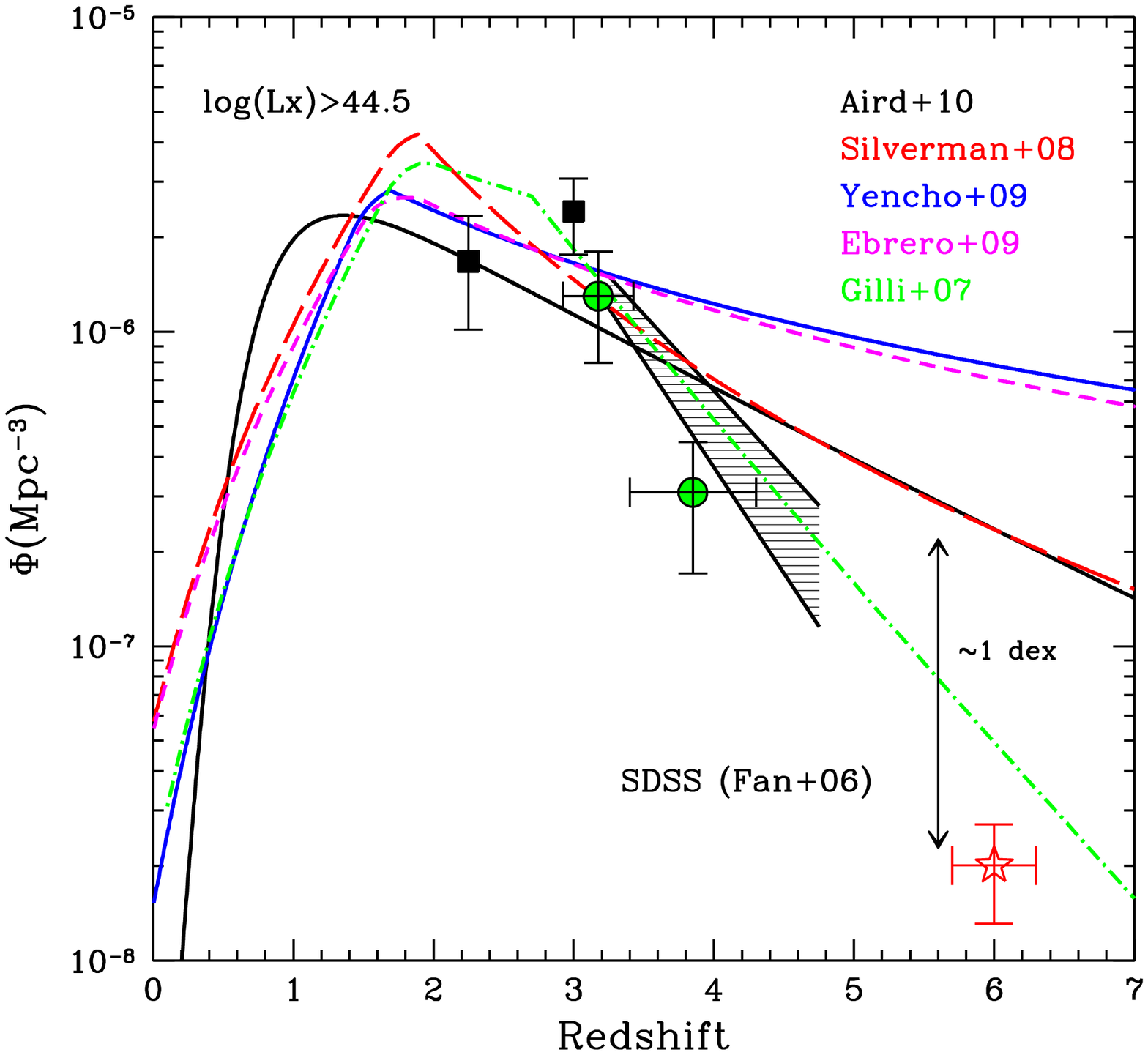}}
\caption{\footnotesize
The number density of log$L_{2-10 keV} >$ 44.5 erg s$^{-1}$ X--ray selected 
AGN vs. redshift as obtained from published LF,
as labeled \citep{gilli07,silverman,ebrero,yencho,aird} 
along with  
%The green dot dashed curve is from the XRB synthesis model 
%of \citet{gilli} assuming an exponential decline at 
%$z>$ 2.7. 
%The space density of X--ray selected AGN 
datapoints from the XMM--COSMOS surveys \citep[green circles,][]{b09a} 
and from \citet[][black squares]{aird}. 
The red star at $z = 6$ represents a conservative estimate of the
z$\sim6$ QSO space density computed from the optical one assuming
no evolution of the $\alpha_{\rm ox}$. 
For comparison (black shaded area) we plot the results for  
very bright QSOs ($M_{1450} < -27$, \citealt{fan}), rescaled by an 
arbitrary factor.}
\label{id_wide}
\end{figure}
%%%% Two columns %%%%%% 

%%%%%%%%%%%%%%%%%%%%%%%%%%%

%%%% Two columns %%%%%% 
%\begin{table}[]
%\caption{High-z AGN in WFXT surveys}
%\label{abun}
%\begin{center}
%\begin{tabular}{lcccc}
%\hline
%\\
%Survey & S$_{lim}$ &  Area & z$>3$ & z$>6$ \\
%\hline
%\\
%Wide  & 4$\times10^{-15}$ cgs & 20.000 deg$^2$ & $1.26\times10^5$ & 500  \\
%Medium  &  5$\times10^{-16}$ cgs & 3.000 deg$^2$ & $2.25\times10^5$ & 1000 \\ 
%Deep  &  3$\times10^{-17}$ cgs & 100 deg$^2$ & 3(6)$\times10^4$ & 300($>300$) \\
%\hline
%\end{tabular}
%\end{center}
%\end{table}
%%%%% Two columns %%%%%% 

\subsection{The likelihood ratio technique}
A statistical, powerful method extensively exploited in deep XMM-Newton
and Chandra surveys in the past years to look for the correct counterparts 
of X--ray sources is the ``likelihood ratio'' 
($LR$) technique \citep{ss,b05}.
The method calculates the probability 
that a source is the correct association
by weighting the information
on the X--ray to optical distance, the surface density of (possible)
false coincidence background objects and the brightness of the 
chosen counterpart: \\
%More specifically, the $LR$ is  defined as the 
%ratio between the probability that the source is the correct identification 
%and the corresponding probability of being a background, unrelated object, 
%i.e.: \\
\centerline{$LR = \frac{q(m) f(r)}{n(m)}$}\\
%where {\it q(m)} is the expected probability distribution, as a
%function of magnitude, of the true counterparts, {\it f(r)} is the 
%probability distribution function of the positional errors of the 
%X--ray sources assumed to be a two--dimensional Gaussian, and {\it n(m)} 
%is the surface density of background objects with magnitude {\it m}. 
The object with the highest LR value\footnote{
{\it q(m)} is the expected probability distribution, as a
function of magnitude, of the true counterparts, {\it f(r)} is the 
probability distribution function of the positional errors of the 
X--ray sources assumed to be a two--dimensional Gaussian, and {\it n(m)} 
is the surface density of background objects with magnitude {\it m}.}  
(above a certain threshold; see
\citealt{ss} for details) is the most likely counterpart;
when two or more sources have comparable LR values, a unique identification
is not possible and both sources have a similar probability
of being the correct identification (``ambiguous" sources). Using catalogs
extracted from different bands (e.g., optical and infrared) may lead to 
different choices of the correct counterparts, and this information should
be taken into account, too. %\citep{luo10}.
In the following we will show the potentiality (and the challenges) 
on the use of the LR technique applied to WFXT data and the multiwavelength
datasets available. We will make the case separately for the 
Wide (F$_{0.5-2} \gsimeq3\times10^{-15}$ erg cm$^{-2}$ s$^{-1}$), 
Medium (F$_{0.5-2} \gsimeq5\times10^{-16}$ erg cm$^{-2}$ s$^{-1}$), and 
Deep (F$_{0.5-2} \gsimeq3\times10^{-17}$ erg cm$^{-2}$ s$^{-1}$)
parts of the WFXT survey (Rosati et al. 2010, this volume), based on the experience 
developed in the framework of the XMM-COSMOS \citep{has}, 
C-COSMOS \citep{elvis} and CDFS \citep{luo08} surveys, 
%the two major programs undertaken today for the study of AGN  demography, 
where {\it multiwavelength} 
catalogs (e.g. optical to mid-infrared) resulted crucial to 
keep the fraction of ambiguous or false identification at minimim.  
%\citep[see][]{b10,luo10}.

%circle depends on the accuracy of the X--ray source positioning and on
%the magnitude of the optical counterpart. 
%At R$>24$ the probability of chance association 

\subsection{Wide and Medium survey: COSMOS lessons} 
To quantify the expected efficiency of the LR technique on the sources
detected in the WFXT Wide survey,
we first limited the XMM-COSMOS sample \citep{cap09} at
fluxes larger than the expected limiting flux of the WFXT Wide survey
(F$_{0.5-2 keV}>3\times10^{-15}$ erg cm$^{-2}$ s$^{-1}$) and comparable
to those expected for the eROSITA deep survey \citep[][see also Cappelluti et al. 2010,
this volume]{erosita}. 
Then we looked at the breakdown of the combined optical and IR 
identifications from the LR technique: 95\% of the sources 
have been provided ``secure" associations, while the remaining 5\% 
show ambiguous counterparts in the \citet{b10} catalog. %
The reliability of the method has been tested a posteriori using Chandra,
and turned out to be 99.6\%: only one of the 245 unique/reliable XMM-COSMOS sources at fluxes 
larger than the WFXT Wide survey resulted associated to the wrong optical 
counterpart. Moreover, the statistical
properties (such as redshifts, magnitudes, colors etc.) of the primary and 
secondary counterpart within the ambiguous sources are almost 
{\it indistinguishable}, and therefore the choice of the counterpart among 
the two does not in principle affect the characterization of the full X--ray 
population.

The WFXT Medium survey has been desgigned to cover $\sim3000$ deg$^{2}$ at
fluxes of the order of $\sim5\times10^{-16}$ erg cm$^{-2}$ s$^{-1}$,
i.e. comparable to the depth reached by the C-COSMOS survey \citep{elvis}.
Following a procedure similar to that applied to XMM-COSMOS data (see
above), and thanks to the smaller ($<1"$) angular resolution of
{\it Chandra} with respect to XMM-{\it Newton}, \citet{civ10} were able to 
provide secure associations for more than 95\% of the sources detected 
above the expected WFXT Medium survey limit. 
The fraction of ambiguous sources in this sample is reduced to $\sim2$\%,
only 1.4\% of the X--ray sources are not identified. 

Taking into account that the WFXT positional accuracy is expected to
be better than that of XMM-{\it Newton} (HEW=$5"-10"$ for WFXT vs. HEW $\sim15"$ for
XMM-{\it Newton}), and only slightly worse than the {\it Chandra} one 
(HEW $\sim2"$ when averaged across the FOV), we can safely conclude that counterpart identification 
would not be an issue for the WFXT Wide and Medium surveys, provided that the depth 
of the optical and IR ancillary data is enough to match the X--ray fluxes 
(see Section 2.4).

%%%% Two columns %%%%%%%
%\begin{figure*}[t!]
%\resizebox{\hsize}{!}{\includegraphics[clip=true]{xid376.ps}
%\includegraphics[clip=true]{cdfs104.ps}}
%\caption{\footnotesize
%{\it First ans second panels}: 
%HST/ACS (left) and Chandra (right) cutouts of the XMM-COSMOS source XID \#376 
%\citep[from][]{b10}. The large circle marks a 3" reference error circle for
%XMM-Newton. The small (0.5") circles mark the positions of the two "ambiguous"
%counterparts.
%{\it Third and fourth panels}: 
%HST/ACS (left) and IRAC 4.5 $\mu$m (right) cutouts of the CDFS source XID 
%\#148 \citep[from][]{luo10} with superimposed a representative Chandra 
%error circle (1.5"); the small (0.5") circles mark the positions of the two 
%"ambiguous" counterparts (see text for details).}
%
%\label{id_wide}
%\end{figure*}
%%%% Two columns %%%%%% 

%%%%%%%%%%%%%%%%%%%%%%%%%%%

%\subsection{Medium survey: C-COSMOS lessons} %
%
%96.6\% secure ID 
%2\% ambigue
%1.4\% not id
%

\subsection{Deep survey: CDFS lessons}
The WFXT deep survey (F$_{0.5-2 keV}>3\times10^{-17}$ erg cm$^{-2}$ 
s$^{-1}$ in the soft X--ray band) 
would be almost as deep as the 
CDFS 2Ms survey \citep{luo08},
over an area that is a factor of $\sim1000$ larger. 
The detailed identification analyses for the 2Ms CDFS sources \citep{luo10}, 
implementing likelihood ratio matching accross five bands (R, z, K, 
3.6$\mu$m and  1.4 GHz) have shown the power of this approach while also 
quantifying the significant challenges in source identification at faint
magnitudes (R$>25$). 
Indeed, it was possible to provide identifications for 96\% of the X--ray 
sources; among them, 90\% have been classified as unique/secure, and 10\% as 
ambiguous. At these deep X--ray fluxes, the statistical properties 
of primary and secondary counterparts between the ambiguous sources 
are often {\it different}. % (see Figure 1 in \citealt{b09b}).
% i.e. in most cases the two sources have very different SEDs 
%(see Figure 2, lower panels): 
%going from the HST to the IRAC image it is clear that a counterpart 
%not present in the optical band shows up at the center of the X--ray 
%error-box. 
%The probability that the optically faint, IR bright source is the correct 
%counterpart is comparable to that of the optically bright source at the edge 
%of the error box.
Moreover, 4\% of the sources, despite the 
excellent quality and depth of the multiwavelength information available,
were not associated to any counterpart, i.e. the correct 
counterpart is most likely fainter than the image depth. 
This exercise shows that the identification of the faintest 
among the WFXT counterparts in the deep survey may be challenging;
%the WFXT capabilities; 
%the situation may be even worse given that the WFXT 
%PSF is expected to be worse than the Chandra one (HEW $\sim5-10"$ vs. HEW 
%$\sim 2"$). 
In this respect, the HEW goal of 5$"$ 
is a crucial requirement to keep  at acceptable values the (already not negligible) 
identifications issues and to fully characterize the multiwavelength 
properties of the X--ray sources at the highest redshifts.

%%%% Two columns %%%%%% 
\begin{table}[]
\caption{Optical and IR ideal coverage depth for WFXT AGN surveys}
\label{abun}
\begin{center}
\begin{tabular}{lcrr}
\hline
\\
Survey & F$^{lim}_{0.5-2}$ & I &  K \\ %& $\langle I \rangle$ & $\langle K \rangle$ \\
  &   cgs&  & \\% &  &  \\ 
\hline
\\
Wide  & 4$\times10^{-15}$  & 23.0 & 21.5  \\ %&  & \\ 
Medium  &  5$\times10^{-16}$ & 25.0 & 23.0 \\ %&  &\\
Deep  &  3$\times10^{-17}$  & 25.5 & 23.5 \\ %&  & \\
\hline
\end{tabular}
\end{center}
\end{table}
%%%% Two columns %%%%%% 

%%%% One column %%%%%% 
\begin{figure*}[!t]
\begin{center}
\resizebox{11cm}{!}{\includegraphics[clip=true]{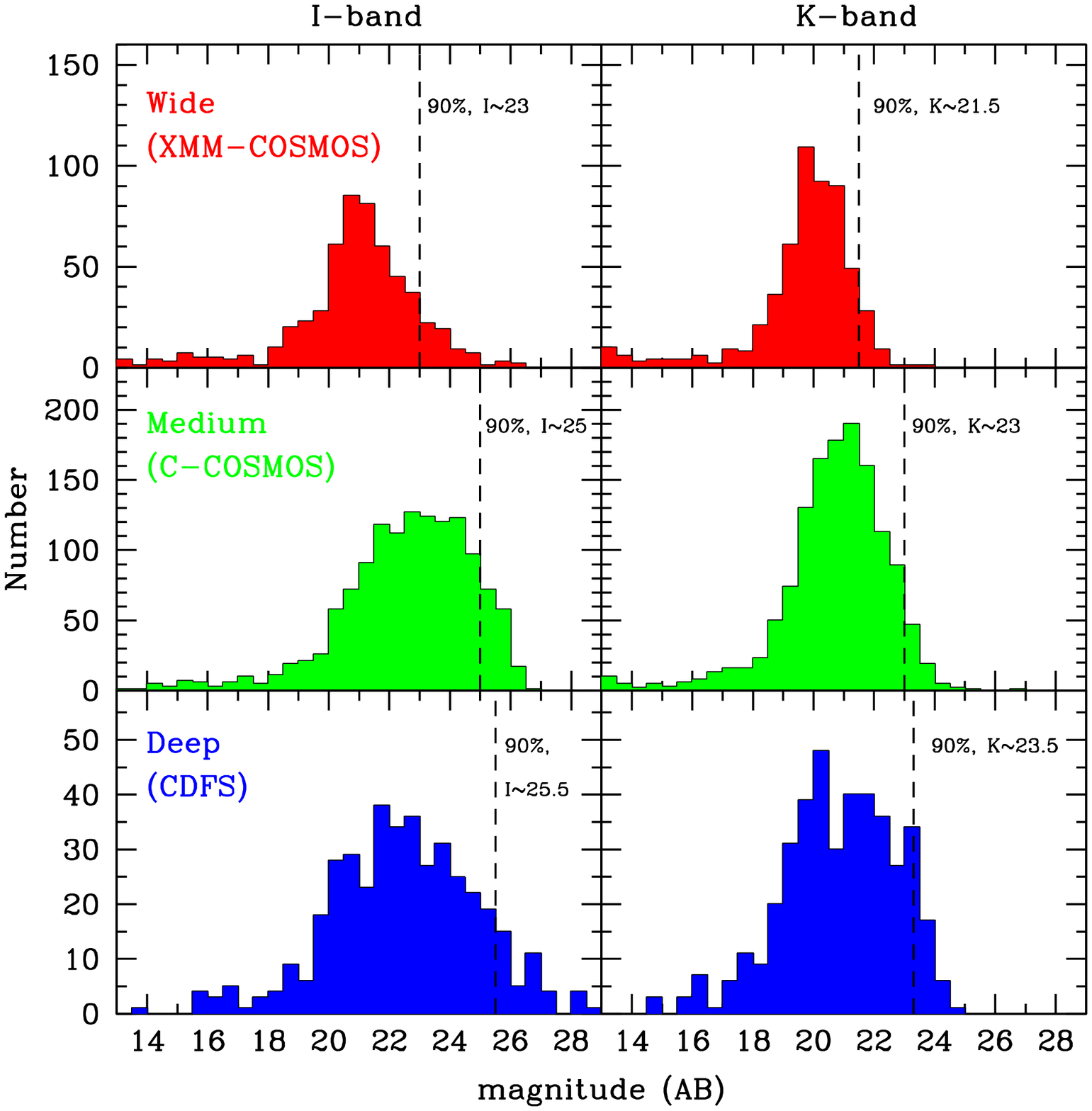}}
%\includegraphic[clip=true]{maghistok}}
\caption{\footnotesize
I-band (left panels) and K-band (right panel) magnitude distributions
expected in the three different WFXT surveys (Wide, Medium and Deep, from
top to bottom). The expected magnitude distributions have been extracted
from the XMM-COSMOS \citep{b10}, C-COSMOS \citep{civ10}, and CDFS \citep{luo10}
samples limited to fluxes F$_{0.5-2}>3\times10^{-15}$ erg cm$^{-2}$ s$^{-1}$,
F$_{0.5-2}>5\times10^{-16}$ erg cm$^{-2}$ s$^{-1}$ and 
F$_{0.5-2}>3\times10^{-17}$ erg cm$^{-2}$ s$^{-1}$ in order to match the
Wide, Medium and Deep limiting fluxes, respectively. 
The dashed lines mark the magnitudes at which most of 
the sources ($90$\%) are identified. }
\label{histomags}
\end{center}
\end{figure*}
%%%% One column %%%%%% 

%%%% One column %%%%%% 
%\begin{figure*}[]
%\resizebox{\hsize}{!}{\includegraphics[clip=true]{104.ps}}
%\caption{
%\footnotesize
%The CDFS source XID \#148 \citep[from][]{luo10}. Left panel: ACS image with superimposed
%the X-ray error circle (2"); middle panel: K-band image; right panel: IRAC image. 
%Going from the HST to the IRAC image it is clear that a counterpart not present
%in the optical band shows up at the center of the error-box. The probability that 
%the optically faint, IR bright source is the correct counterpart is comparable
%to that of the optically bright source at the edge of the error box}
%\label{id_deep}
%\end{figure*}
%%%% One column %%%%%% 

\subsection{Depth of optical infrared images}
The power of the LR technique described in the previous subsections is
strongly related to the depth of the optical and infrared images and catalogs 
that will be used to identify the X-ray sources. 
The challenge will be to provide a homogeneous and (enough) deep coverage 
for the different WFXT surveys.  
At the limiting flux of the WFXT wide survey an optical coverage to I$\sim23$
and K$\sim21.5$ would be enough to identify $\sim90$\% of the X--ray sources
(see Figure 2, upper panels, and Table 1), 
but this should be on the {\it entire} surveyed area. At the time WFXT will
be launched, PanSTARRS\footnote{http://pan-starrs.ifa.hawaii.edu} will have surveyed $\sim30.000$ deg$^2$ 
to I$\sim 24.2$, and will provide imaging 
in at least 5 bands, needed to characterize the SED of the X--ray sources
and isolate high-z candidates (see next Section). 
On a longer timescale, Euclid\footnote{http://sci.esa.int/euclid}
will cover the entire extragalactic sky in the IR down to H$\sim24$ (roughly 
corresponding to K$\sim 23$), and will provide 
also spectra. The LOw Frequency ARray \citep[LOFAR,][]{morganti}, 
that will survey the northern sky down to a flux of 0.8 mJy at 120 MHz
\citep[see Fig. 2 in][]{morganti}, may 
be crucial to correctly identify radio emitters X--ray sources 
(radio AGN and starbursts). 

PanSTARRS will also provide identification for
a substantial fraction ($>50$\%) of the sources detected in the WFXT Medium and Deep surveys.
%i.e. only the brightest among the X--ray source population detected at
%the faintest fluxes.  
In order to identify a fraction as large as 90\% of the sources in these surveys, 
a coverage in the optical and  near--infrared down to I$\sim25.5$ and K$\sim23.5$ 
is needed (see Figure 2, middle and lower panels, and Table 1).
LSST %\footnote{http://www.lsst.org/lsst} 
is a proposed facility 
expected to cover the southern sky down to  I$\sim 27$ \citep{lsst};
similarly to PanSTARRS, LSST will also provide multiband photometry
at a depth comparable to the I-band limit. 
The coordination with present and next generation facilities is mandatory, 
in order to choose the areas for the deep surveys which maximize the 
availability of the deepest multiwavelength coverage, in particular:  
JWST\footnote{http://www.jwst.nasa.gov}, 
the PanSTARRS deep survey (I$\sim 28$ over 28 deg$^2$), 
the LSST deep survey (I$\sim 28$ on a few hundreds deg$^2$), 
Euclid (K$\sim 25.5$ on 40 deg$^2$), the VISTA VIDEO survey (down to K=23.5 
over 15 deg$^2$).

\section{Selecting z$>3$ (or z$>6$) AGN}
Photometric redshifts of X--ray selected faint sources (R$=24-27$) are essential for 
enabling science analyses and planning deep spectroscopy, and resulted {\it crucial} 
in isolating the high-z population in, e.g., XMM-COSMOS and CDFS.  
%which constitute
%only a tiny fraction of the total X--ray population. As an example, 
%only 40 objects among 1650 sources ($\sim2$\%) turned out to be associated 
%to z$>3$ QSO \citep{b09a}. 
A similar, detailed source characterization requiring {\it multiband} imaging
may be feasible only for small samples of WFXT sources. In this context, key resources will be again 
the upcoming LSST, PanSTARRS, Euclid, and (as far as spectroscopy for the 
Wide survey is concerned) the SDSSIII-BOSS\footnote{http://www.sdss3.org/boss.php} project.  
However, the high-z population shows on average fainter
magnitudes than the overall X--ray source population (see Figure 3 in \citealt{b09a}),
and therefore may remain among the unidentified population, if deep enough optical
and near infrared coverage is not provided over the full area 
%{\it N.B.: are depth and number of bands enough to get photoz?}. 
Another possibility is to search for X--ray counterparts on preselected
high-z QSO on the basis of optical colours and/or dropouts techniques
\citep[e.g.][]{casey,steidel}, extended including the near-infrared bands
in order to sample the z$>6$ population \citep[e.g.][]{willott}.
In this respect, the unprecedent combination of depth and area of WFXT
will result in a much better characterization
of the physical properties (such as bolometric luminosity and accretion
rate) of the first accreting supermassive black holes. 
Moreover, z$>6$ color selections suffer from significant contamination 
stellar objects (brown dwarfs are overwhelmingly more abundant and the spectroscopy success rate for z$>6$ QSOs 
is only $\sim 20$\%). The complete SED characterization from NIR to 
X--ray will be able to resolve issues on contamination and completeness. 
For a non negligible fraction of the high-z candidates (a few out of a few 
hundreds, see also Matt et al. 2010, this volume), redshifts may be directly 
measured from the FeK$\alpha$ line \citep[see examples in][]{com}.

\section{Conclusions}

\begin{itemize} 

\item[$\bullet$] WFXT will provide orders of magnitudes {\it larger} 
samples of high-redshift (z$>6$) AGN compared to current (e.g. SDSS) 
optical surveys; 

\item[$\bullet$] the counterpart identification for WFXT sources selected
in the Wide survey will be relatively easy,
{\it if} synergies with present and future large area / all sky facilities
(e.g. PanSTARRs, LSST, Euclid) are pursued; 

\item[$\bullet$] the secure identification of the counterparts detected 
in the WFXT Medium and Deep surveys would greatly benefit of 
the {\it smallest possible angular resolution} (the 5" HEW goal is really
auspicable) and should heavily rely in the coordination with the future
optical and NIR deep survey area (e.g. LSST, JWST);

\item[$\bullet$] multiwavelength information is {\it mandatory} in order
to get the redshift and the physical properties of the high-z AGN in the WFXT surveys.

\end{itemize}

\begin{acknowledgements}
We gratefully acknowledge the essential contribution from the COSMOS and 
CDFS teams, and in particular Bin Luo. RG acknoweledges support from the
ASI grant I/088/06/00.
\end{acknowledgements}

\bibliographystyle{aa}

\end{document}